\documentclass[twocolumn]{aastex61}
\usepackage{graphicx}
\usepackage{tablefootnote}

\received{October 9, 2017}
\revised{November 30, 2017}
\accepted{November 30, 2017}
\submitjournal{ApJ}

\shorttitle{Chemical Abundances for HD~23193 and HD~170920}
\shortauthors{K{\i}l{\i}\c{c}o\u{g}lu et al.}
\begin{document}

\title{Behavior of Abundances in Chemically Peculiar Dwarf and Subgiant A-Type Stars: HD~23193 and HD~170920\footnote{Based on observations made at the T\"{U}B{\.I}TAK National Observatory(Program ID 14BRTT150-671), and the Ankara University Observatory, Turkey.}}

\correspondingauthor{\c{S}eyma \c{C}al{\i}\c{s}kan}
\email{seyma.caliskan@science.ankara.edu.tr}

\author{Tolgahan K{\i}l{\i}\c{c}o\u{g}lu}
\affiliation{Ankara University, Science Faculty, Department of Astronomy and Space Sciences, 06100, Ankara, Turkey}

\author{\c{S}eyma \c{C}al{\i}\c{s}kan}
\affil{Ankara University, Science Faculty, Department of Astronomy and Space Sciences, 06100, Ankara, Turkey}

\author{K\"{u}bra\"{o}zge \"{ U}nal}
\affiliation{Ankara University, Science Faculty, Department of Astronomy and Space Sciences, 06100, Ankara, Turkey}

%% Note that the \and command from previous versions of AASTeX is now
%% depreciated in this version as it is no longer necessary. AASTeX 
%% automatically takes care of all commas and "and"s between authors names.

%% AASTeX 6.1 has the new \collaboration and \nocollaboration commands to
%% provide the collaboration status of a group of authors. These commands 
%% can be used either before or after the list of corresponding authors. The
%% argument for \collaboration is the collaboration identifier. Authors are
%% encouraged to surround collaboration identifiers with ()s. The 
%% \nocollaboration command takes no argument and exists to indicate that
%% the nearby authors are not part of surrounding collaborations.

%% Mark off the abstract in the ``abstract'' environment. 
\begin{abstract}

To understand the origin of the abundance peculiarities of non-magnetic A-type stars, we present the first detailed chemical abundance analysis of a metallic line star HD~23193 (A2m) and an A-type subgiant HD~170920 (A5) which could have been a HgMn star on the main sequence. Our analysis is based on medium (R$\sim$14\,000) and high (R$\sim$40\,000) resolution spectroscopic data of the stars. The abundance of 18 elements are derived: C, O, Na, Mg, Al, Si, S, Ca, Sc, Ti, Cr, Mn, Fe, Ni, Zn, Sr, Y, and Ba. The masses of HD~23193 and HD~170920 are estimated from evolutionary tracks, as 2.3$\pm 0.1$~$M_{\odot}$ and 2.9$\pm 0.1$~$M_{\odot}$. The ages are found 635$\pm$33~Myr for HD~23193 and 480$\pm$50~Myr for HD~170920 using isochrones. The abundance pattern of HD~23193 shows deviations from solar values in the iron-peak elements and indicates remarkable overabundances of Sr (1.16), Y (1.03), and Ba (1.24) with respect to the solar abundances. We compare the derived abundances of this moderately rotating ($v$sin$i = 37.5~$km\,s$^{-1}$) Am star to the theoretical chemical evolution models including rotational mixing. The theoretically predicted abundances resemble our derived abundance pattern, except for a few elements (Si and Cr). For HD~170920, we find nearly solar abundances, except for C ($-0.43$), S (0.16), Ti (0.15), Ni (0.16), Zn (0.41), Y (0.57), and Ba (0.97). Its low rotational velocity ($v$sin$i = 14.5~$km\,s$^{-1}$), reduced carbon abundance, and enhanced heavy element abundances suggest that the star is most-likely an evolved HgMn star. 

\end{abstract}

%% Keywords should appear after the \end{abstract} command. 
%% See the online documentation for the full list of available subject
%% keywords and the rules for their use.
\keywords{stars: abundances --- stars: chemically peculiar --- stars: individual (HD~23193, HD~170920)}
%% From the front matter, we move on to the body of the paper.
%% Sections are demarcated by \section and \subsection, respectively.
%% Observe the use of the LaTeX \label
%% command after the \subsection to give a symbolic KEY to the
%% subsection for cross-referencing in a \ref command.
%% You can use LaTeX's \ref and \label commands to keep track of
%% cross-references to sections, equations, tables, and figures.
%% That way, if you change the order of any elements, LaTeX will
%% automatically renumber them.

%% We recommend that authors also use the natbib \citep
%% and \citet commands to identify citations.  The citations are
%% tied to the reference list via symbolic KEYs. The KEY corresponds
%% to the KEY in the \bibitem in the reference list below. 

\section{Introduction} \label{sec:intro}

Superficially normal A-type stars may exhibit differences in elemental abundances with respect to the Sun in considerable ranges \citep{adelmanetal15}. Most of these differences can be seen in the heavy element abundances \citep{adelmanunsuree07}. Also, a group of non-magnetic A-type stars have remarkable peculiar abundances. These metallic line A-type (hereafter Am) stars are usually characterized by underabundances of some of the light elements (i.e., He, C, O, Ca, and Sc), slight/moderate overabundances of some of the iron-peak elements (Fe, Cr, Ni, etc.), and remarkable overabundances of many heavy elements such as Sr, Y, Zr, and Ba. They are divided into subclasses according to their observed properties, as classical, marginal (or mild), and hot Am stars (for details, see \citealt{kurtz78}). However, there is no clear boundaries between these subclasses nor between normal and metallic A-type stars.

\citet{michaud70} suggested diffusion processes as a mechanism for most of the observed chemical peculiarities in the atmospheres of A-type stars. He established that if the atmosphere of a star is not disturbed by convection nor any other mixing process (such as turbulence and meridional circulation) in a shorter time than the diffusion time scale (e.g., order of $10^{4}$ years for Mn~III), certain elements can move toward the top of the atmosphere by large radiative forces, causing overabundances. Diffusion processes were then studied in detail for many elements by \citet{michaudetal76} and \citet[including turbulent motions]{vauclairetal78}. Detailed quantitative evolution models, including 28 chemical elements and turbulent transport, were calculated by \citet{richeretal00}. Despite the fact that their theoretically predicted surface abundances generally agree with those of observed Am/Fm stars, they concluded that an additional mixing mechanism was needed to explain observations. The effect of rotational mixing and mass loss on chemical abundances were also modeled/discussed in \citet{talon06} and \citet{vick10}. The abundance analysis of Am stars with various $v$sin$i$ may help to reveal how stellar rotation affects diffusion process. 

HD~23193 is classified as an A2m: by \citet{cowley68} and \citet{cowleyetal69}, A3~III by \citet{osawa59}, and A4p-Ba enhanced by \citet{burwell38}. \citet{floquet70} states that HD~170920 might be an Am star, and its spectral type is in the range of A3-A7. \citet{henry71} perform the measurements of Ca~II~K line strength for 223 stars, and the authors include the 369 field stars having k-index measurements. In the study, HD~23193 is listed as an Am star with a spectral type of A2. 

Both stars are present in the \textit{Catalogue of Am stars with known spectral types} given by \citet{hauck73}. In this catalog, the spectral classes A7 (from metallic lines) and A3 (from Ca~II~K line) are given for HD~170920, and A2 (from Ca~II~K line) for HD~23193. Both HD~23193 and HD~170920 is available in \textit{General Catalogue of Ap and Am stars} \citep{rensonetal91}. In the catalog, a doubtful peculiarity is noted only for HD~23193. Two stars also are included in the study of \citet{rensonmanfroid09}, with no any remarks on their peculiarity. The rotational and radial velocities of the stars are collected and listed in Table~\ref{tab1}. 

\begin{table}
\caption{Collected rotational and radial velocities of the target stars.} 
\centering
\begin{tabular}{cccc}
\hline\hline
    ~        & HD~23193    &HD~170920      & References\\
\hline
 $v$sin$i$    & 33          &13             & \citet{abtmorrell95} \\ 
 (km\,s$^{-1}$)      & 42          &21             & \citet{royeretal02}\\ 
\hline
$v_{\rm r}$ & 21.8        &$-$27.3        & \citet{duflotetal95} \\
(km\,s$^{-1}$)      & 27.3$\pm$0.6&$-$27.3$\pm$2.9& \citet{gontcharov06} \\
\hline
\end{tabular} \label{tab1}
\end{table}

\citet{mcdonaldetal12} calculated the effective temperatures and luminosities of the stars by comparing BT-SETTL model atmospheres to spectral energy distributions generated from infrared photometric data. According to their results, HD~23193 has an effective temperature of 8400~K, and a luminosity of 37.33~L$_{\odot}$. The fundamental parameters of HD~170920 are 7497~K, and 113.21~L$_{\odot}$. The distances 90.58~pc for HD~23193 and 188.32~pc for HD~170920 were also given in \citet{mcdonaldetal12}. The effective temperature of 8960~K was estimated for HD~23193 by \citet{glagolevskij94} using ($B2-V1$)-$T_{\rm eff}$ calibration of \citet{haucknorth93}, based on the optical photometry.

We present the chemical abundances of moderately rotating ($v$sin$i = 37.5~$km\,s$^{-1}$) Am star HD~23193 and a slowly rotating ($v$sin$i = 14.5~$km\,s$^{-1}$) A type subgiant HD~170920, to understand the origin of their abundance pattern. In Section~\ref{sec:obs}, the observations and data reduction methods are given. The details of the abundance analysis are explained in Section~\ref{sec:analysis}. We present the evolutionary status of the stars in Section ~\ref{sec:evo}, and it is followed with the abundance results Section~\ref{sec:aburesu}. Finally, we discuss and summarize the results of the study in Section~\ref{sec:conc}.     

\section{Observation and data reduction} \label{sec:obs}

HD~23193 ($\alpha$[2000]=$03^{h}44^{m}31^{s}$; $\delta$[2000]=+36$^{\circ}$27' 36'') and HD~170920 ($\alpha$[2000]=$18^{h}31^{m}57^{s}$; $\delta$[2000]=$-01^{\circ}$00' 11'') are bright stars of V$\sim$5.59 and 5.94 magnitudes. The optical region spectra of the stars were obtained both T\"{U}B\.{I}TAK National Observatory (TUG) with 1.5~m telescope and Ankara University Kreiken Observatory (AUKR) with 0.4~m telescope. The TUG spectra with high-resolution (R$\sim$40,000) and high-SNR ($>$150) were acquired by Coude Echelle Spectrograph, covering a wavelength range of 3900 to 7500~\AA\ with an exposure time of 3600~s on October 13, 2016. The AUKR spectra with medium resolution (R$\sim$14,000) and high-SNR ($\sim$200) were obtained by Shelyak eShel Spectrograph, covering a wavelength range of 4340 to 7400~\AA\ with an exposure time of 3600~s on September 24, 2017. The medium resolution spectra were only used to analyze $H_\mathrm{\beta}$ lines (for details see Section~\ref{sec:atmpar}), since the narrow ($\sim$60~\AA) echelle orders of high resolution spectra do not cover any hydrogen Balmer lines entirely. 
The calibration images (bias, dark, flat-fielding, arc spectra) were also taken for each night. We performed the data reduction, wavelength calibration, and normalization of the high-resolution spectra with standard IRAF\footnote{IRAF is distributed by the National Optical Astronomy Observatories, which are operated by the Association of Universities for Research in Astronomy, Inc., under cooperative agreement with the National Science Foundation.} (Image Reduction and Analysis Facility) packages, as described in the study of \citet{caliskanetal15}. For the reduction of the medium resolution echelle spectra, we used the AUKR data reduction pipeline based on the conventional reduction steps\footnote{\url{http://www.shelyak.com/dossier.php?id_dossier=47}}.

\section{Abundance analysis} \label{sec:analysis}

\subsection{Atmospheric parameters} \label{sec:atmpar}
\begin{figure}
\begin{center}
\includegraphics[width=\columnwidth]{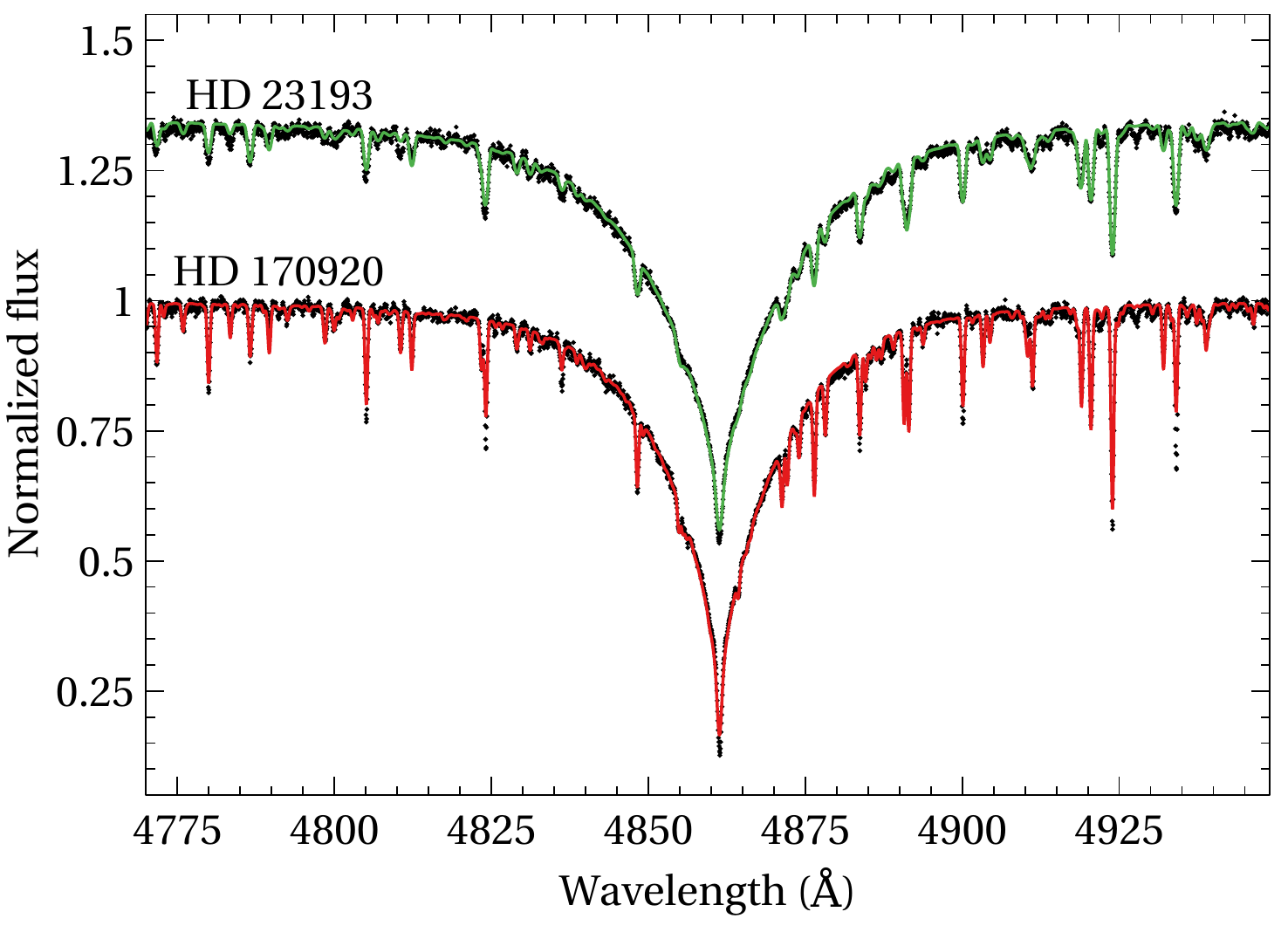}
\caption{Comparison between the observed (black points) and synthetic (colored solid lines) H$_{\beta}$ profiles of the stars for the adopted parameters given in Table 2.}
\label{Fig1}
\end{center}
\end{figure}

\begin{figure}
\begin{center}
\includegraphics[width=\columnwidth]{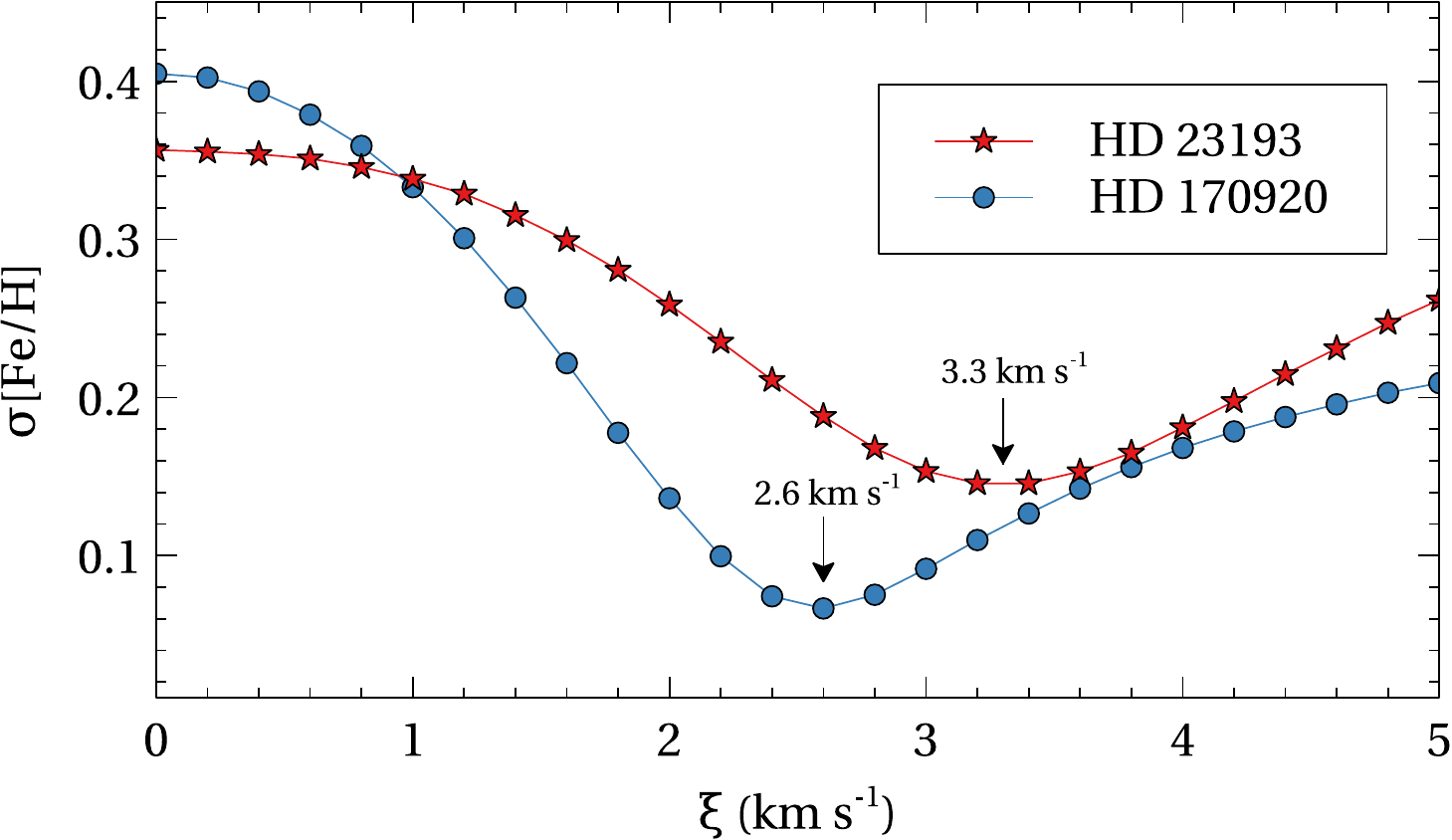}
\caption{Determination of the microturbulences by minimizing the standard deviation
of $\rm [Fe/H]$.}
\label{Fig2}
\end{center}
\end{figure}

As a first step for the abundance analysis of HD~23193 and HD~170920, the atmospheric parameters ($T_{\rm eff}$ and log~$g$) were determined from Str{\"o}mgren, Geneva, and Johnson (only for $T_{\rm eff}$) photometric data, using the calibrations of \citet{napiwotzki97}, \citet{kunzlietal97}, and \citet{flower96}. Using the mean values of these atmospheric parameters, the initial model atmospheres were computed by ATLAS9 \citep{kurucz93, sbordoneetal04}, which assumes the line formation in LTE, plane-parallel geometry. No convection was used for the model of HD~23193, whereas a convection with a mixing length parameter of 0.5 was used for HD~170920. We then produced the synthetic spectra using SYNSPEC49 \citep{hubenylanz95} and its SYNPLOT interface to refine the $T_{\rm eff}$ and log~$g$ values of the stars from their observed H$_{\beta}$ line profiles, which are sensitive both $T_{\rm eff}$ and log~$g$. In Fig.~\ref{Fig1}, we show the observed and synthetic H$_{\beta}$ profile fitting for both stars. The rotational velocities of the stars were derived by comparing the observed iron lines with the synthetic iron lines calculated for various $v$sin$i$ values. The derived rotational velocities are in agreement with the average of the velocities given in Table~\ref{tab1}. In order to derive the microturbulent velocity of the stars, we derived the iron abundance $\rm [Fe/H]$ by using 40 unblended Fe~II lines for a set of microturbulent velocities ranging from 0.0 to 5.0 km\,s$^{-1}$. Fig.~\ref{Fig2} shows the standard deviation of the derived $\rm [Fe/H]$ as a function of the microturbulent velocity. The adopted microturbulent velocity is the value which minimizes the standard deviation. The photometric $T_{\rm eff}$ and log~$g$ pairs and final adopted $T_{\rm eff}$, log~$g$, $\xi$, and $v$sin$i$ of the stars are tabulated in Table~\ref{tab2}.

\begin{table*}
\caption{Derived atmospheric parameters and rotational velocities of the stars.} 
\centering
\begin{tabular}{cccccccccc}
\hline\hline
 ~&\multicolumn{5}{c}{Photometric}& \multicolumn{4}{c}{Adopted from Spectroscopy} \\
\cline{2-6}\cline{7-10} \\
 ~&\multicolumn{2}{c}{Str{\"o}mgren\tablenotemark{a}}&\multicolumn{2}{c}{Geneva\tablenotemark{b}} &\multicolumn{1}{c}{Johnson\tablenotemark{c}}& & & &\\
\hline 
    Star     & $T_{\rm eff}$&log~$g$& $T_{\rm eff}$&log~$g$& $T_{\rm eff}$&$T_{\rm eff}$&log~$g$& $\xi$& $v$sin$i$\\
~            & (K) & (in cgs) &  (K) & (in cgs) & (K) &(K) & (in cgs) & (km\,s$^{-1}$) & (km\,s$^{-1}$)\\
\hline            
HD~23193 &8593 &3.78 &8892 &4.16 &8727  & $8800\pm100$ & $3.80\pm0.10$ & 3.3& $37.5\pm1.2$ \\ 
HD~170920&8052 &3.15 &8108$^d$ &3.12 &8670  & $8250\pm150$ & $3.15\pm0.10$ & 2.6& $14.5\pm0.8$ \\
\hline
\end{tabular} \label{tab2}
\medskip
\tablenotetext{a}{The data are from \citet{paunzen15} for HD~23193, and \citet{hauckmermilliod98} for HD~170920.}
\tablenotetext{b}{The data are from \citet{hauckcurchod80}.}
\tablenotetext{c}{The data are from \citet{hogetal00}.}
\tablenotetext{d}{$E(B-V) = 0.07$ for HD~170920.}

\end{table*}

\subsection{Chemical abundances from spectrum synthesis} \label{sec:cabu}

For the spectrum synthesis, we used the atomic line list of \citet{kilicogluetal16} compiled from Kurucz's Line Database\footnote{\url{http://kurucz.harvard.edu/linelists/gfhyperall/gfhyperall.dat}},  VALD \citep{piskunovetal95, ryabchikovaetal97, kupkaetal99, kupkaetal00}, NIST atomic database \citep{NIST_ASD}, and various available sources of the oscillator strengths. We initially selected unblended (or slightly blended) lines of the elements (141 lines for HD~23193, 281 lines for HD~170920) in the high resolution spectra of the stars. We then fitted these selected lines with synthetic line profiles produced by SYNSPEC49 and its SYNPLOT interface. The code was modified to minimize the $\chi^{2}$ between the model and the observed points, using Levenberg-Marquardt algorithm \citep{markwardt09}. 

The abundances of the elements were adjusted until the best-fit between the synthetic and observed line profiles. The synthetic spectra were broadened by the rotational velocity, microturbulence, and instrumental profile. We computed ATLAS9 model atmosphere to derive the abundances of both stars. For the chemically peculiar Am star HD~23193, we also computed an ATLAS12 \citep{kurucz05} model atmosphere based on the derived abundances with ATLAS9, and refined its elemental abundances. 

\section{Evolutionary Status} \label{sec:evo}

\begin{figure}
\begin{center}
\includegraphics[width=\columnwidth]{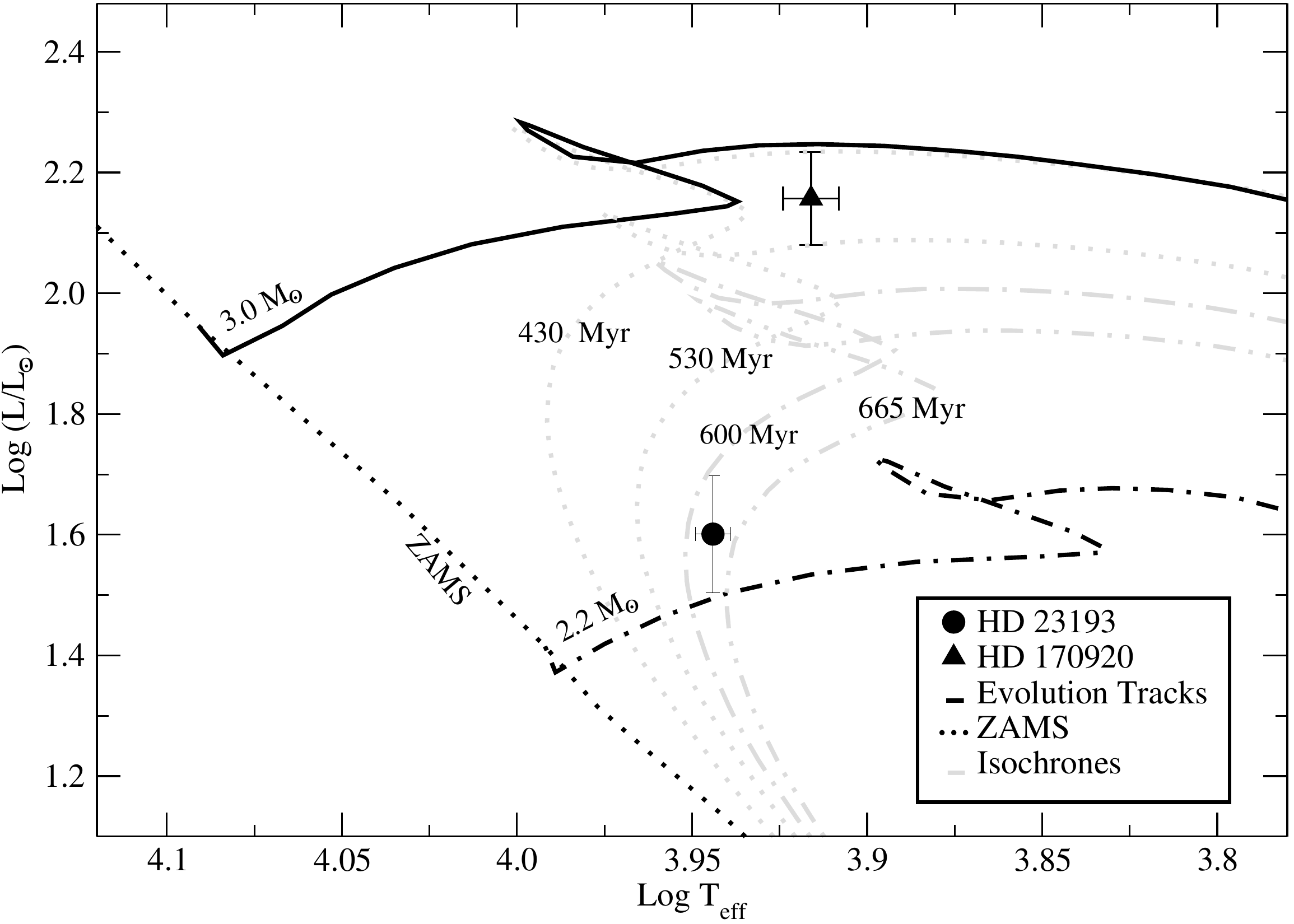}
\caption{Location of the stars on the HR diagram.}
\label{Fig3}
\end{center}
\end{figure}

Using the effective temperatures (derived in this study) and the luminosities (calculated from the parameters given in Table~\ref{tab3}), we plotted the target stars on a theoretical HR diagram (Fig. \ref{Fig3}). We used the evolutionary tracks having various masses (i.e., 2.2, 3.0\,$M_{\odot}$) and solar metallicity to estimate the masses of the stars \citep{salasnichetal00}. For their ages, we considered the isochrones with various ages (i.e., 430, 530, 600, 665\,Myr) and solar metallicity \citep{bressanetal12}. We found a mass of 2.3$\pm$0.1~$M_{\odot}$ for HD~23193, and 2.9$\pm$0.1~$M_{\odot}$ for HD~170920. The estimated age of HD~23193 is 635$\pm$33~Myr, and the age of HD~170920 is 480$\pm$50~Myr. 

\begin{table*}
\caption{Calculated parameters of the stars for H-R diagram.}
\centering
\begin{tabular}{ccccccc} 
\hline\hline
Star&$m_{\rm v}$ & $\pi$ &$M_{\rm v}$ & BC & log$(L/L_{\odot})$ & log $T_{\rm eff}$ \\
  ~ & (mag)      & (mas) & (mag)      & (mag) & ~               & (K)                \\
\hline%
HD~23193 & 5.59&11.04$\pm$0.31&   0.80$\pm$0.01&$-$0.085&1.60$\pm$0.10&3.944$\pm$0.005 \\
HD~170920& 5.94&5.31 $\pm$0.45&$-$0.66$\pm$0.02&$-$0.014&2.16$\pm$0.08&3.916$\pm$0.008 \\
\hline
\end{tabular}\label{tab3}
\tablecomments{$m_{\rm v}$, $\pi$, and BC are from \citet{hogetal00}, \citet{vanleeuwen07}, and\\ \citet{torres10}, respectively.}
\end{table*}

As shown in Fig. \ref{Fig3}, the age (635$\pm$33~Myr) and mass (2.3$\pm$0.1~$M_{\odot}$) of HD~23193 shows that the star located between ZAMS\footnote{Zero age main sequence} and TAMS\footnote{Terminal age main sequence}. However, the position of HD~170920 with $M$=2.9$\pm$0.1~$M_{\odot}$ and $\tau$=480$\pm$50~Myr indicates that the star just left from the main sequence and it is moving into giant region. It should consequently be called a subgiant.

\section{Results} \label{sec:aburesu}

The derived elemental abundances for HD~23193 and HD~170920 are given in Table~\ref{tab4}. We have indicated the abundances as nearly solar, whenever they are in the range of $-0.10$$<$[X/H]$<$+0.10. The values larger than 0.10 are considered as slightly overabundant and larger than 0.25 ones are overabundant with respect to the solar. Similar statements has also been used for underabundances (i.e., slightly underabundant: $-$0.25$<$[X/H]$<$$-$0.10, underabundant: [X/H]$<$$-$0.25). All abundances are given with respect to the solar values, taken from \citet{grevessesauval98}. The abundance errors for elements with only one detectable line have been adopted as 0.2. 

\begin{table}
\caption{Derived chemical abundances with standard deviations ($\sigma$) for HD~23193 and HD~170920.}
\centering
\begin{tabular}{crccrcc}
\hline\hline
    ~     &\multicolumn{3}{c}{HD~23193}&\multicolumn{3}{c}{HD~170920}\\
\hline   
Elements& $\rm{[X/H]}$&$\sigma$&N& $\rm{[X/H]}$&$\sigma$&N\\ 
\hline
C &$-$0.43 & 0.20  & 1  & $-$0.13	&	0.13	&	17	\\
O &$-$0.07 & 0.20  & 1  & $-$0.08	&	0.07	&	8	\\
Na&   0.10 & 0.10  & 2  & $-$0.04   &   0.01    &   2   \\
Mg&$-$0.06 & 0.15  & 5  &    0.08	&	0.30	&	4	\\
Al&   0.55 & 0.20  & 1  &    0.00	&	0.20    &	1	\\
Si&   0.35 & 0.17  & 3  & $-$0.01	&	0.19	&	3	\\
S &   0.21 & 0.20  & 1  &    0.16	&	0.13	&	4	\\
Ca&$-$0.19 & 0.24  & 5  & $-$0.04	&	0.14	&	24	\\
Sc&$-$0.10 & 0.04  & 4  & $-$0.06	&	0.11	&	11	\\
Ti&   0.19 & 0.17  & 16 &    0.15	&	0.22	&	28	\\
Cr&   0.13 & 0.18  & 11 &    0.01	&	0.14	&	24	\\
Mn&    $-$ &  $-$  & $-$&    0.03   &   0.06    &   3   \\
Fe&   0.27 & 0.16  & 76 &    0.00	&	0.07	&	40	\\
Ni&   0.42 & 0.15  & 6  &    0.16	&	0.10	&	11	\\
Zn&   0.68 & 0.20  & 1  &    0.41	&	0.15	&	2	\\
Sr&   1.16 & 0.20  & 1  &     $-$   &   $-$     & $-$	\\
Y &   1.03 & 0.23  & 2  &    0.57	&	0.11	&	5	\\
Ba&   1.24 & 0.33  & 4  &    0.97	&	0.38	&	4	\\
\hline
\end{tabular}\label{tab4}
\tablecomments{N is number of lines. The abundances of \\ HD~23193 were derived from ATLAS12.}
\end{table}

For HD~170920, all elements are nearly solar except for Carbon which is slightly underabundant, S-Ti-Ni which are slightly overabundant, and Zn-Y-Ba which are overabundant. For HD~23193, we found overabundances in Al-Si-Fe-Ni-Zn-Sr-Y-Ba, slightly overabundances in S-Ti-Cr, slightly underabundances in Ca-Sc, and underabundance in C while the other elements are nearly solar. We note that the abundance of C-O-Al-S-Zn-Sr for HD~23193 and Al for HD 170920 are derived from only one line and the values must be taken with caution. The Na abundance of both stars are derived from Na I lines at 5682 and 5688 \AA. The Na~I lines at 5890 and 5895 \AA\ were not used due to the strong non-LTE effects and interstellar absorption \citep{takedaetal09}.

\section{Discussion and Conclusion} \label{sec:conc}

\begin{figure*}
\begin{center}
\includegraphics[width=\textwidth]{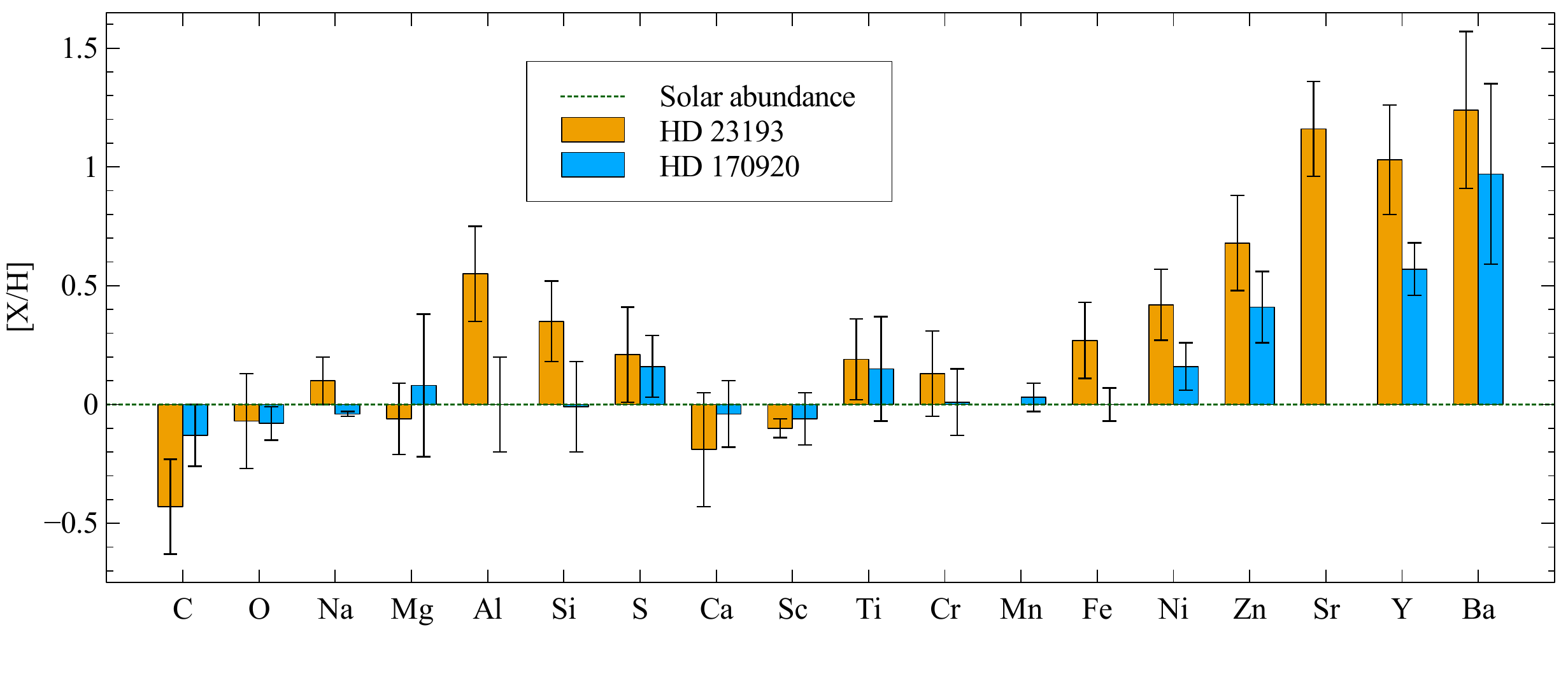}
\caption{Abundance pattern of HD~23193 and HD~170920. The error bars indicate the standard deviation of abundances derived from individual lines.}
 \label{Fig4}
\end{center}
\end{figure*}

We derived the abundances of 18 elements from the high resolution spectra of HD~23193 and HD~170920. The abundance pattern of both stars are shown in Fig.~\ref{Fig4}. Also, we estimated their masses and ages. The mass and age of HD~23193 are 2.3$\pm$0.1~$M_{\odot}$ and 635$\pm$33~Myr. For HD~170920, the predicted mass and age are 2.9$\pm$0.1~$M_{\odot}$ and 480$\pm$50~Myr.

%The abundances of HD~23193 are compatible with those of Am-type stars. This type of stars generally exhibit underabundances in the light elements, slightly overabundance in the iron-peak elements, and highly overabundance in the rare-earth elements relative to the solar. They may also exhibit underabundances in C, O, Ca and/or Sc. 
In the pattern of HD~23193, the abundance of the elements heavier than Mn roughly tend to increase with atomic number while C, Ca and Sc elements are slightly underabundant. The chemical abundance analysis of the star thus confirms its Am characteristic. This star (with $v$sin$i$=37.5~km\,s$^{-1}$) is one of the good examples of Am stars having relatively higher rotational velocities.

The detailed abundance analysis of HD~170920 have revealed that C and O are marginally underabundant, S-Ti-Ni are slightly overabundant, Zn-Y-Ba are overabundant and the other elements do not significantly deviate from the solar abundances. Considering the main-sequence position of the stars with $2.9~M_{\odot}$, this star originally comes from the $T_{\rm eff}$-$\mathrm{log}(L/L_\odot)$ domain where HgMn stars are found (e.g., primary component of $\chi$ Lupi, \citealt{lebouquinetal13}) . The low rotational velocity, reduced light element abundances, and enhanced heavy metal content suggest that this star is most-likely an evolved HgMn star. The present abundance pattern thus represents residual peculiarities when it reaches to the subgiant domain.

We qualitatively compare the derived abundance pattern of the Am star HD~23193 (with 2.3~$M_{\odot}$, and 635~Myr) with that of predicted from Montreal code by \citet[Fig. 12]{talon06} for a mass of 2.5~$M_{\odot}$. As predicted by the model at 600~Myr, we indeed derived sub-solar abundances for C-Ca, and super-solar abundances for Al-Ti-Fe-Ni in the atmosphere of HD~23193. However, the star does not show a remarkable overabundance for Cr, and it shows an overabundance for Si, which are in contradiction with the model prediction indicating nearly solar values. On the other hand, both elements Cr and Si are quite solar in the atmosphere of the A-type subgiant HD~170920. The theoretical abundances predicted by the model for a rotational velocity of 50~km\,s$^{-1}$ (denoted as 2.5P1 and 2.5P3 in their Fig. 12) are in better agreement with the observed abundances of HD~23193 than the model with a rotational velocity of 15~km\,s$^{-1}$ (denoted as 2.5P2). We thus conclude that the reduced overabundance of iron peak elements in the atmosphere of HD~23193 might be the result of its slightly high $v$sin$i$ (37.5 km\,s$^{-1}$). In order to understand how stellar rotation reduces the effects of diffusion process, the abundance analysis of more Am stars with various $v$sin$i$ are needed.

%The pattern is in agreement with normal A-type stars, as seen in the study of \citet{adelmanunsuree07}. We conclude that HD~170920 is a chemically normal star, performing its first detailed abundance analysis. Due to its slow rotational velocity (14.5 km\,s$^{-1}$) and $\rm [Fe/H]$ of 0.0, HD~170920 is a good candidate of abundance standard for A-type stars. It is also worth to mention that, HD~170920 has many flat bottomed lines belonging to C~I, Sc~II, Ti~II, Cr~II, Fe~I, Fe~II species in its spectrum. These flat bottomed lines might indicate that the star is a pole-on star, as in Vega. Its equatorial velocity can be much larger than its $v$sin$i$.

\begin{acknowledgements}
The authors acknowledge the T\"UB\.ITAK National Observatory (TUG) with the project ID 14BRTT150-671, and the support by the research fund of Ankara University (BAP) with the project ID 15A0759001. We thank D. \"Ozt\"urk for her help during the observations at AUKR. We also thank the anonymous reviewer for his/her valuable suggestions and highly appreciate the constructive comments on the origin of the chemical peculiarity of HD 170920, which significantly improved the quality of the paper.  
\end{acknowledgements}

\bibliographystyle{aasjournal}
\bibliography{kilicoglu.bib}

\begin{thebibliography}{}
\expandafter\ifx\csname natexlab\endcsname\relax\def\natexlab#1{#1}\fi
\providecommand{\url}[1]{\href{#1}{#1}}

\bibitem[{{Abt} \& {Morrell}(1995)}]{abtmorrell95}
{Abt}, H.~A., \& {Morrell}, N.~I. 1995, \apjs, 99, 135

\bibitem[{{Adelman} {et~al.}(2015){Adelman}, {Gulliver}, \&
  {Kaewkornmaung}}]{adelmanetal15}
{Adelman}, S.~J., {Gulliver}, A.~F., \& {Kaewkornmaung}, P. 2015, \pasp, 127,
  340

\bibitem[{{Adelman} \& {Unsuree}(2007)}]{adelmanunsuree07}
{Adelman}, S.~J., \& {Unsuree}, N. 2007, Baltic Astronomy, 16, 183

\bibitem[{{Bressan} {et~al.}(2012){Bressan}, {Marigo}, {Girardi}, {Salasnich},
  {Dal Cero}, {Rubele}, \& {Nanni}}]{bressanetal12}
{Bressan}, A., {Marigo}, P., {Girardi}, L., {et~al.} 2012, \mnras, 427, 127

\bibitem[{{Burwell}(1938)}]{burwell38}
{Burwell}, C.~G. 1938, \apj, 88, 278

\bibitem[{\c{C}al{\i}{\c s}kan {et~al.}(2015)\c{C}al{\i}{\c s}kan,
  {K{\i}l{\i}{\c c}o{\u g}lu}, {Elmasl{\i}}, {Nasolo}, {Avc{\i}}, \&
  {Albayrak}}]{caliskanetal15}
\c{C}al{\i}{\c s}kan, {\c S}., {K{\i}l{\i}{\c c}o{\u g}lu}, T., {Elmasl{\i}},
  A., {et~al.} 2015, New A, 34, 6

\bibitem[{{Cowley} {et~al.}(1969){Cowley}, {Cowley}, {Jaschek}, \&
  {Jaschek}}]{cowleyetal69}
{Cowley}, A., {Cowley}, C., {Jaschek}, M., \& {Jaschek}, C. 1969, \aj, 74, 375

\bibitem[{{Cowley}(1968)}]{cowley68}
{Cowley}, A.~P. 1968, \pasp, 80, 453

\bibitem[{{Duflot} {et~al.}(1995){Duflot}, {Figon}, \&
  {Meyssonnier}}]{duflotetal95}
{Duflot}, M., {Figon}, P., \& {Meyssonnier}, N. 1995, \aaps, 114, 269

\bibitem[{{Floquet}(1970)}]{floquet70}
{Floquet}, M. 1970, \aaps, 1, 1

\bibitem[{{Flower}(1996)}]{flower96}
{Flower}, P.~J. 1996, \apj, 469, 355

\bibitem[{{Glagolevskij}(1994)}]{glagolevskij94}
{Glagolevskij}, Y.~V. 1994, Bulletin of the Special Astrophysics Observatory,
  38, 152

\bibitem[{{Gontcharov}(2006)}]{gontcharov06}
{Gontcharov}, G.~A. 2006, Astronomy Letters, 32, 759

\bibitem[{{Grevesse} \& {Sauval}(1998)}]{grevessesauval98}
{Grevesse}, N., \& {Sauval}, A.~J. 1998, \ssr, 85, 161

\bibitem[{{Hauck}(1973)}]{hauck73}
{Hauck}, B. 1973, \aaps, 10, 385

\bibitem[{{Hauck} \& {Curchod}(1980)}]{hauckcurchod80}
{Hauck}, B., \& {Curchod}, A. 1980, \aap, 92, 289

\bibitem[{{Hauck} \& {Mermilliod}(1998)}]{hauckmermilliod98}
{Hauck}, B., \& {Mermilliod}, M. 1998, \aaps, 129, 431

\bibitem[{{Hauck} \& {North}(1993)}]{haucknorth93}
{Hauck}, B., \& {North}, P. 1993, \aap, 269, 403

\bibitem[{{Henry} \& {Hesser}(1971)}]{henry71}
{Henry}, R.~C., \& {Hesser}, J.~E. 1971, \apjs, 23, 421

\bibitem[{{H{\o}g} {et~al.}(2000){H{\o}g}, {Fabricius}, {Makarov}, {Urban},
  {Corbin}, {Wycoff}, {Bastian}, {Schwekendiek}, \& {Wicenec}}]{hogetal00}
{H{\o}g}, E., {Fabricius}, C., {Makarov}, V.~V., {et~al.} 2000, \aap, 355, L27

\bibitem[{{Hubeny} \& {Lanz}(1995)}]{hubenylanz95}
{Hubeny}, I., \& {Lanz}, T. 1995, \apj, 439, 875

\bibitem[{{K{\i}l{\i}{\c c}o{\u g}lu} {et~al.}(2016){K{\i}l{\i}{\c c}o{\u
  g}lu}, {Monier}, {Richer}, {Fossati}, \& {Albayrak}}]{kilicogluetal16}
{K{\i}l{\i}{\c c}o{\u g}lu}, T., {Monier}, R., {Richer}, J., {Fossati}, L., \&
  {Albayrak}, B. 2016, \aj, 151, 49

\bibitem[{Kramida {et~al.}(2015)Kramida, {Yu.~Ralchenko}, Reader, \& {and NIST
  ASD Team}}]{NIST_ASD}
Kramida, A., {Yu.~Ralchenko}, Reader, J., \& {and NIST ASD Team}. 2015, {NIST
  Atomic Spectra Database (ver. 5.3), [Online]. Available:
  {\tt{http://physics.nist.gov/asd}} [2017, October 1]. National Institute of
  Standards and Technology, Gaithersburg, MD.}, ,

\bibitem[{{Kunzli} {et~al.}(1997){Kunzli}, {North}, {Kurucz}, \&
  {Nicolet}}]{kunzlietal97}
{Kunzli}, M., {North}, P., {Kurucz}, R.~L., \& {Nicolet}, B. 1997, \aaps, 122,
  51

\bibitem[{{Kupka} {et~al.}(1999){Kupka}, {Piskunov}, {Ryabchikova}, {Stempels},
  \& {Weiss}}]{kupkaetal99}
{Kupka}, F., {Piskunov}, N., {Ryabchikova}, T.~A., {Stempels}, H.~C., \&
  {Weiss}, W.~W. 1999, \aaps, 138, 119

\bibitem[{{Kupka} {et~al.}(2000){Kupka}, {Ryabchikova}, {Piskunov}, {Stempels},
  \& {Weiss}}]{kupkaetal00}
{Kupka}, F.~G., {Ryabchikova}, T.~A., {Piskunov}, N.~E., {Stempels}, H.~C., \&
  {Weiss}, W.~W. 2000, Baltic Astronomy, 9, 590

\bibitem[{{Kurtz}(1978)}]{kurtz78}
{Kurtz}, D.~W. 1978, \apj, 221, 869

\bibitem[{{Kurucz}(1993)}]{kurucz93}
{Kurucz}, R. 1993, ATLAS9 Stellar Atmosphere Programs and 2 km/s grid.~Kurucz
  CD-ROM No.~13.~ Cambridge, Mass.: Smithsonian Astrophysical Observatory,
  1993., 13

\bibitem[{{Kurucz}(2005)}]{kurucz05}
{Kurucz}, R.~L. 2005, Memorie della Societa Astronomica Italiana Supplementi,
  8, 14

\bibitem[{{Le Bouquin} {et~al.}(2013){Le Bouquin}, {Beust}, {Duvert}, {Berger},
  {M{\'e}nard}, \& {Zins}}]{lebouquinetal13}
{Le Bouquin}, J.-B., {Beust}, H., {Duvert}, G., {et~al.} 2013, \aap, 551, A121

\bibitem[{{Markwardt}(2009)}]{markwardt09}
{Markwardt}, C.~B. 2009, in Astronomical Society of the Pacific Conference
  Series, Vol. 411, Astronomical Data Analysis Software and Systems XVIII, ed.
  D.~A. {Bohlender}, D.~{Durand}, \& P.~{Dowler}, 251

\bibitem[{{McDonald} {et~al.}(2012){McDonald}, {White}, {Zijlstra}, {Guzman
  Ramirez}, {Szyszka}, {van Loon}, {Lagadec}, \& {Jones}}]{mcdonaldetal12}
{McDonald}, I., {White}, J.~R., {Zijlstra}, A.~A., {et~al.} 2012, \mnras, 427,
  2647

\bibitem[{{Michaud}(1970)}]{michaud70}
{Michaud}, G. 1970, \apj, 160, 641

\bibitem[{{Michaud} {et~al.}(1976){Michaud}, {Charland}, {Vauclair}, \&
  {Vauclair}}]{michaudetal76}
{Michaud}, G., {Charland}, Y., {Vauclair}, S., \& {Vauclair}, G. 1976, \apj,
  210, 447

\bibitem[{{Napiwotzki}(1997)}]{napiwotzki97}
{Napiwotzki}, R. 1997, \aap, 322, 256

\bibitem[{{Osawa}(1959)}]{osawa59}
{Osawa}, K. 1959, \apj, 130, 159

\bibitem[{{Paunzen}(2015)}]{paunzen15}
{Paunzen}, E. 2015, \aap, 580, A23

\bibitem[{{Piskunov} {et~al.}(1995){Piskunov}, {Kupka}, {Ryabchikova}, {Weiss},
  \& {Jeffery}}]{piskunovetal95}
{Piskunov}, N.~E., {Kupka}, F., {Ryabchikova}, T.~A., {Weiss}, W.~W., \&
  {Jeffery}, C.~S. 1995, \aaps, 112, 525

\bibitem[{{Renson} {et~al.}(1991){Renson}, {Gerbaldi}, \&
  {Catalano}}]{rensonetal91}
{Renson}, P., {Gerbaldi}, M., \& {Catalano}, F.~A. 1991, \aaps, 89, 429

\bibitem[{{Renson} \& {Manfroid}(2009)}]{rensonmanfroid09}
{Renson}, P., \& {Manfroid}, J. 2009, \aap, 498, 961

\bibitem[{{Richer} {et~al.}(2000){Richer}, {Michaud}, \&
  {Turcotte}}]{richeretal00}
{Richer}, J., {Michaud}, G., \& {Turcotte}, S. 2000, \apj, 529, 338

\bibitem[{{Royer} {et~al.}(2002){Royer}, {Grenier}, {Baylac}, {G{\'o}mez}, \&
  {Zorec}}]{royeretal02}
{Royer}, F., {Grenier}, S., {Baylac}, M.-O., {G{\'o}mez}, A.~E., \& {Zorec}, J.
  2002, \aap, 393, 897

\bibitem[{{Ryabchikova} {et~al.}(1997){Ryabchikova}, {Piskunov}, {Kupka}, \&
  {Weiss}}]{ryabchikovaetal97}
{Ryabchikova}, T.~A., {Piskunov}, N.~E., {Kupka}, F., \& {Weiss}, W.~W. 1997,
  Baltic Astronomy, 6, 244

\bibitem[{{Salasnich} {et~al.}(2000){Salasnich}, {Girardi}, {Weiss}, \&
  {Chiosi}}]{salasnichetal00}
{Salasnich}, B., {Girardi}, L., {Weiss}, A., \& {Chiosi}, C. 2000, \aap, 361,
  1023

\bibitem[{{Sbordone} {et~al.}(2004){Sbordone}, {Bonifacio}, {Castelli}, \&
  {Kurucz}}]{sbordoneetal04}
{Sbordone}, L., {Bonifacio}, P., {Castelli}, F., \& {Kurucz}, R.~L. 2004,
  Memorie della Societa Astronomica Italiana Supplementi, 5, 93

\bibitem[{{Takeda} {et~al.}(2009){Takeda}, {Kang}, {Han}, {Lee}, \&
  {Kim}}]{takedaetal09}
{Takeda}, Y., {Kang}, D.-I., {Han}, I., {Lee}, B.-C., \& {Kim}, K.-M. 2009,
  \pasj, 61, 1165

\bibitem[{{Talon} {et~al.}(2006){Talon}, {Richard}, \& {Michaud}}]{talon06}
{Talon}, S., {Richard}, O., \& {Michaud}, G. 2006, \apj, 645, 634

\bibitem[{{Torres}(2010)}]{torres10}
{Torres}, G. 2010, \aj, 140, 1158

\bibitem[{{van Leeuwen}(2007)}]{vanleeuwen07}
{van Leeuwen}, F. 2007, \aap, 474, 653

\bibitem[{{Vauclair} {et~al.}(1978){Vauclair}, {Vauclair}, \&
  {Michaud}}]{vauclairetal78}
{Vauclair}, G., {Vauclair}, S., \& {Michaud}, G. 1978, \apj, 223, 920

\bibitem[{{Vick} {et~al.}(2010){Vick}, {Michaud}, {Richer}, \&
  {Richard}}]{vick10}
{Vick}, M., {Michaud}, G., {Richer}, J., \& {Richard}, O. 2010, \aap, 521, A62

\end{thebibliography}

\clearpage

\end{document}